\documentclass[aps,pra]{revtex4-2}

\usepackage{color}

\usepackage{graphicx,amsmath,amssymb}

\begin{document}

\title{Simplified scheme for continuous-variable entanglement distillation:  multicopy distillation of Gaussian entanglement without heralding Gaussian measurements}

\author{Jaromír Fiurášek}
\affiliation{Department of Optics, Faculty of Science, Palack\'y University, 17.\ listopadu 12, 77900  Olomouc, Czech Republic}

\begin{abstract}
Entanglement of continuous-variable Gaussian  states can be distilled by combination of de-Gaussifying operation such as single-photon subtraction and iterative heralded Gaussification. 
Here we present and analyze a simplified equivalent version of such entanglement distillation protocol, where the Gaussian measurements utilized in heralded Gaussification are eliminated and are absorbed into the preparation of suitable input Gaussian states of the simplified protocol. The simplified scheme contains less detectors and  its overall success probability increases in comparison with the original scheme, while producing completely equivalent outputs. Our simplification of the entanglement distillation  protocol closely parallels the recently proposed simplification of a scheme for breeding optical single-mode Gottesman-Kitaev-Preskill states [H. Aghaee Rad \emph{et al.},  Nature \textbf{638}, 912 (2025)]. We investigate operation of the simplified entanglement distillation scheme for both pure and mixed input states and clarify how multicopy  distillation of Gaussian entanglement emerges in a setup without any heralding Gaussian measurements. 

\end{abstract}

\maketitle

\section{Introduction}

\begin{figure*}[!b!]
\centerline{\includegraphics[width=0.75\linewidth]{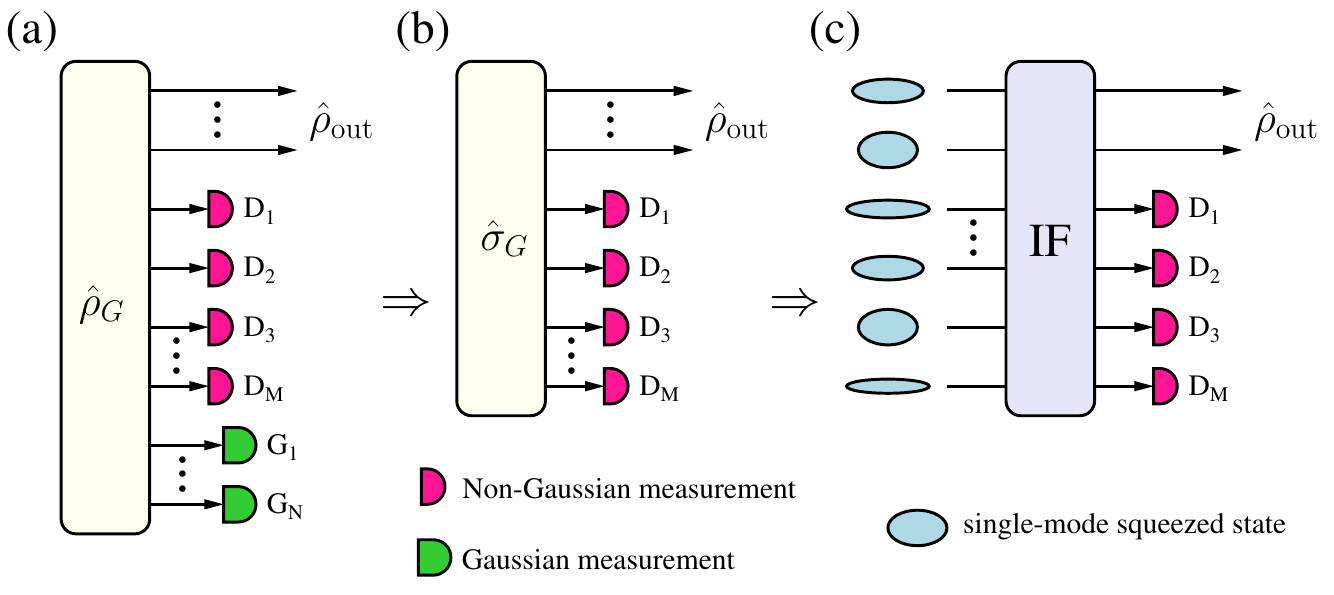}}
\caption{Equivalence between quantum-state preparation schemes with and without Gaussian measurements. In (a), a general multimode Gaussian state $\hat{\rho}_G$ is prepared, and some of the modes are measured with Gaussian or non-Gaussian detectors to generate the output state $\hat{\rho}_{\mathrm{out}}$.  Conditioning on specific outcomes of Gaussian measurements can be interpreted as a preparation of a different input Gaussian state $\hat{\sigma}_G$ from $\hat{\rho}_G$, which yields the simplified scheme in  (b). (c) For pure states, Bloch-Messiah decomposition can be applied to generate the input Gaussian state from pure single-mode squeezed states by interference in a multimode interferometer.}
\label{figgeneral}
\end{figure*}

Generation of nonclassical states of light is essential for optical quantum information processing. A large class of optical quantum-state engineering protocols is based on generation of multimode squeezed Gaussian states followed by heralding measurements on the auxiliary output modes \cite{Su2019,Lvovsky2020,Walschaers2021,Biagi2022,Larsen2025,Aralov2025,Hanamura2025}. This includes, among others, various schemes based on conditional photon subtraction \cite{Dakna1997,Wenger2004,Ourjoumtsev2006,Nielsen2006,Wakui2007,Ourjoumtsev2007nonlocal,Takahashi2008,Ourjoumtsev2009,Nielsen2010,Huang2015,Asavanant2017,Endo2023,Endo2025}, conditional photon addition \cite{Zavatta2004,Barbieri2010,Kumar2013,Jeong2014,Fadrny2024,Chen2024}, and measurements on one part of two-mode squeezed vacuum state \cite{Paris2003,Lvovsky2001,Yukawa2013,Cooper2013,Bouillard2019,Kawasaki2022}. The heralding measurements can be both non-Gaussian and Gaussian and many schemes combine these two types of measurements \cite{Ourjoumtsev2007,Etesse2015,Sychev2017,Cotte2022,Simon2024,Konno2024}. Non-Gaussian measurements usually involve photon counting represented by projection onto Fock state $|n\rangle$ with $n>0$, or conditioning on click of a binary on/off detector that can distinguish the presence and absence of photons. The Gaussian measurements include projection onto vacuum or projection onto an eigenstate of some quadrature operator, accomplished with homodyne detector. These latter measurements are typically employed in the state breeding schemes where elementary non-Gaussian states are combined and merged together to prepare a more complex state that better approximates the targeted resource such as the Gottesman-Kitaev-Preskill (GKP) states  \cite{Vasconcelos2010,Konno2024} or cat-like states formed by superposition of two coherent states \cite{Etesse2015,Sychev2017}. Gaussian measurements also play essential role in iterative Gaussification protocols  which asymptotically convert input non-Gaussian quantum states into Gaussian ones \cite{Browne2003,Eisert2004,Campbell2012,Dostal2024}. These Gaussification protocols can be utilized for distillation of continuous variable entanglement \cite{Browne2003,Eisert2004,Hage2008,Hage2010,Fiurasek2010} and also for distillation and enhancement of quadrature squeezing \cite{Franzen2006,Grebien2022}.

Every measurement increases the complexity of the resulting experimental setup. It turns out that the  state preparation schemes depicted in Fig.~\ref{figgeneral}(a) can be converted into equivalent simpler schemes without Gaussian measurements, see Fig.~\ref{figgeneral}(b). Recently, this has been explicitly pointed out in Ref. \cite{Rad2025} in the context of iterative schemes for breeding GKP states and it was shown that such  simplification gives rise to schemes for generation of GKP states that are closely related to Gaussian Boson sampling \cite{Lund2014,Hamilton2017,Paesani2019,Zhong2020,Zhong2021,Thekkadath2022}. Importantly, the resulting simplified setup  involves only conditioning on detection of nonzero number of photons in each auxliary mode \cite{Rad2025,Larsen2025,Takase2023}. Simplification of the state-preparation scheme becomes possible when the Gaussian and non-Gaussian measurements in the original scheme are performed in parallel and the scheme does not include any feedforward. One can then imagine that the Gaussian measurements in Fig.~\ref{figgeneral}(a) are performed before the non-Gaussian measurements. The Gaussian measurements are then performed on a multimode Gaussian state and the resulting output state is again Gaussian. 

If the state $\hat{\sigma}_G$  is pure, we can furthermore apply   the Bloch-Messiah decomposition \cite{Braunstein2005}.  According to this decomposition, any pure multimode Gaussian state can be generated from input product single-mode squeezed vacuum states with suitable squeezing constants $r_j$ by their interference at multimode passive optical interferometer, followed by coherent displacements of each output mode. We are often interested in schemes where the Gaussian states are not initially displaced and one conditions on projections onto vacuum or onto the quadrature eigenstate with quadrature value $0$. In such cases the coherent displacements vanish and the scheme simplifies further.

In Ref. \cite{Rad2025} the  simplification of quantum-state breeding scheme was  performed for a setup dedicated to generation of approximate single-mode GKP states. In the present work we extend and apply this concept to iterative Gaussification of two-mode entangled states that were previously de-Gaussified by local conditional single-photon subtractions. We show that the Gaussian measurements present in the iterative Gaussification can indeed be avoided and we present an equivalent protocol that involves only non-Gaussian measurements, namely projections of the auxiliary modes onto single-photon Fock states $|1\rangle$. Notably, this alternative but equivalent entanglement distillation protocol can converge to a Gaussian state although it does not involve any Gaussian measurements.
To obtain physical insight we first consider pure-state version of the protocol but then we take losses into account. We find that, importantly, the LOCC structure of the CV entanglement distillation protocol remains unchanged by the elimination of the Gaussian measurements even in the presence of losses.

The rest of the paper is organized as follows. In Sec. II we review continuous-variable entanglement distillation protocol based on conditional photon subtractions and heralded Gaussification. In Sec. III we derive an equivalent two-copy entanglement distillation scheme that does not include any Gaussian measurements. Functioning  of this scheme for pure input states is discussed in Sec. IV. The simplified scheme is compared with alternative scheme based on generalized photon subtraction in Sec. V.  Mixed input states are considered in Sec. VI. Simplified entanglement distillation protocol for arbitrary number of input copies is presented in Sec. VII. Finally, Sec. VIII contains a brief discussion and conclusions.

\section{Entanglement distillation by iterative Gaussification}

In this paper we consider continuous-variable entanglement distillation protocol based on distribution of entangled Gaussian states, de-Gaussification by local conditional single-photon photon subtractions \cite{Opatrny2000,Takahashi2010,Kurochkin2014,Dirmeier2020} and subsequent iterative Gaussification \cite{Browne2003,Eisert2004,Hage2008}. Note that the de-Gaussification is a crucial part of the protocol because Gaussian entanglement cannot be distilled by local Gaussian operations and classical communication \cite{Eisert2002,Giedke2002,Fiurasek2002}. 

\begin{figure*}
\centerline{\includegraphics[width=\linewidth]{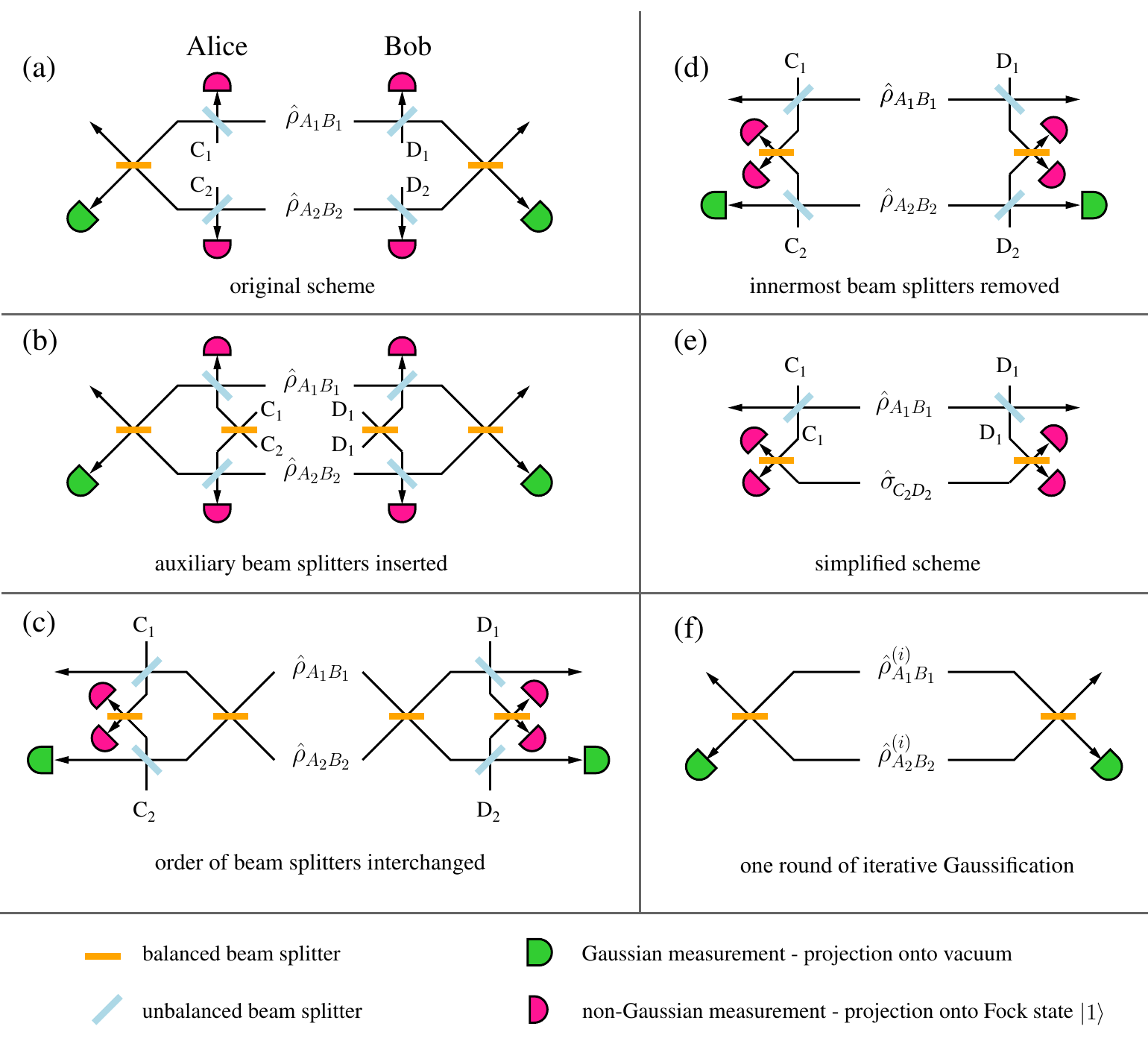}}
\caption{Construction of simplified equivalent scheme for single iteration of the entanglement distillation protocol. Orange bars denote balanced beam splitters, light blue bars represent unbalanced beam splitters that originally serve for conditional photon subtraction. Green detectors are projections onto vacuum while the pink detectors indicate projections onto the single-photon states $|1\rangle$.  Both $\hat{\rho}$ and $\hat{\sigma}$ in panels (a)-(e) denote Gaussian states with vanishing coherent displacements. The auxiliary input modes  C$_j$ and $D_j$ in panels (a)-(d) are prepared in vacuum state. Panel (f) shows one round of iterative Gaussification protocol that probabilistically  maps two copies of a non-Gaussian state $\hat{\rho}^{(i)}$ onto a single copy of a partially Gaussified state.}
\label{figsingle}
\end{figure*}

Imagine a scheme where copies of pure two-mode suqeezed vacuum state are generated at a central source S and are distributed over two lossy channels with transmittance $\eta$ to Alice and Bob, who thus possess many copies of a mixed Gaussian state $\hat{\rho}_{AB}$.  Alice and Bob jointly de-Gaussify the state by applying local single-photon subtractions \cite{Opatrny2000,Olivares2003,Takahashi2010,Kurochkin2014,Kitagawa2006,DellAnno2013,Yang2009,Zhang2010,Navarrete2012,Wang2015,Bartley2013}. This is accomplished by mixing locally each mode with  a vacuum state at an unbalanced beam splitter with transmittance T, followed by projection of the auxiliary output mode onto the single-photon state $|1\rangle$, c.f. Fig.~\ref{figsingle}(a). 
Subsequently, Alice and Bob  Gaussify their shared state by iteratively applying the procedure depicted in Fig.~\ref{figsingle}(a). Two copies of the state are locally interfered at balanced beam splitters and the auxiliary output mode on each side is projected onto vacuum, which heralds succesful Gaussification step. As shown in Ref. \cite{Eisert2004}, if the initial state satisfies certain conditions, the iterative heralded Gaussification converges to a Gaussian state. Formally, a single round of the iterative heralded Gaussification protocol can be described by the following nonlinear map \cite{Eisert2004},
\begin{widetext}
\begin{equation}
\hat{\rho}_{A_1B_1}^{(i+1)}={\mathcal{N}_i^{-1}}  _{A_2B_2}\! \langle 0,0|  \left(\hat{U}_{A_1A_2}\otimes \hat{U}_{B_1B_2}\right)\left(\hat{\rho}_{A_1B_1}^{(i)}\otimes \hat{\rho}_{A_2B_2}^{(i)}\right)\left( \hat{U}_{A_1A_2}^\dagger\otimes \hat{U}_{B_1B_2}^\dagger\right) |0,0\rangle_{A_2B_2},
\label{Gaussificationmap}
\end{equation}
where
\begin{equation}
\mathcal{N}_i =\mathrm{Tr}_{A_1B_1} \left[ _{A_2B_2}\! \langle 0,0|  \left(\hat{U}_{A_1A_2}\otimes \hat{U}_{B_1B_2}\right)\left(\hat{\rho}_{A_1B_1}^{(i)}\otimes \hat{\rho}_{A_2B_2}^{(i)}\right)\left( \hat{U}_{A_1A_2}^\dagger\otimes \hat{U}_{B_1B_2}^\dagger\right) |0,0\rangle_{A_2B_2}\right],
\end{equation}
\end{widetext}
and $\hat{U}$ denotes a two-mode unitary operation describing interference at a balanced beam splitter,
\begin{equation}
\hat{U}_{AB}=\exp\left[ \frac{\pi}{4}\left( \hat{a}^\dagger\hat{b}-\hat{a}\hat{b}^\dagger \right)\right].
\label{UBSAB}
\end{equation}
Here and in what follows, the subscripts indicate the modes the operators act on. 

To obtain some insight, it is instructive to neglect losses and consider input pure two-mode squeezed vacuum state (TMSV)
\begin{equation}
|\psi_{\mathrm{TMSV}}(\lambda)\rangle=\sqrt{1-\lambda^2}\sum_{n=0}^\infty \lambda^n |n,n\rangle,
\label{TMSV}
\end{equation}
with squeezing parameter $\lambda=\tanh r$. Conditional  subtraction of  $m$ photons from mode A is described by a nonunitary operator 
\begin{equation}
\hat{K}_{m}=\frac{(1-T)^{m/2}}{\sqrt{m!}} T^{\hat{n}_A/2} \hat{a}^m
\label{Km}
\end{equation}
 where $\hat{a}$ denotes the annihilation operator associated with mode A and $\hat{n}_A=\hat{a}^\dagger \hat{a}$ is the corresponding single-mode photon number operator. When a single photon is subtracted from each mode of the state (\ref{TMSV}), we obtain a non-normalized conditional state \cite{Kitagawa2006}
\begin{widetext}
\begin{equation}
\hat{K}_{1,A}\otimes\hat{K}_{1,B}|\psi_{\mathrm{TMSV}}(\lambda)\rangle =\sqrt{1-\lambda^2}\lambda(1-T)\sum_{n=0}^\infty (n+1)(T\lambda)^n |n,n\rangle.
\label{TMSVsubtracted}
\end{equation}
\end{widetext}
Depending on the values of $T$ and $\lambda$, the photon-subtracted state (\ref{TMSVsubtracted}) can exhibit increased entanglement and higher two-mode squeezing in comparison to the input Gaussian  two-mode squeezed vacuum state (\ref{TMSV}). Therefore, the photon subtraction can be used to improve the performance of the Braunstein-Kimble continuous-variable quantum teleportation scheme \cite{Opatrny2000,Olivares2003,Yang2009,Wang2015} and of the continuous-variable entanglement swapping \cite{Wang2024swapping}. At the same time, the state (\ref{TMSVsubtracted}) is non-Gaussian and its Wigner function is negative at some  regions of the phase space. It is even in principle possible to observe violation of Bell inequalities with homodyne detection when using the photon-subtracted state (\ref{TMSVsubtracted}) as a resource \cite{Nha2004,Patron2004}.

Subsequent iterative Gaussification (\ref{Gaussificationmap}) drives the state (\ref{TMSVsubtracted}) to a Gaussian two-mode squeezed vacuum state with squeezing parameter $\lambda_D=2T\lambda$, provided that $|\lambda_D|<1$. Otherwise, the protocol diverges. As shown in Ref. \cite{Eisert2004}, the convergence of the protocol depends on the  amplitudes of the lowest Fock states up to total photon number $2$. More specifically, the ratio of the amplitudes of states $|11\rangle$ and $|00\rangle$  in the superposition (\ref{TMSVsubtracted}) determines  the squeezing parameter $\lambda_D$ of the asymptotically distilled two-mode squeezed vacuum. For $T>1/2$, the protocol enhances the squeezing and entanglement of the distilled pure two-mode squeezed vacuum state. 

In practice, we are often interested in distillation of mixed states distributed over lossy quantum channels. The considered distillation protocol can increase the squeezing and entanglement of  mixed input Gaussian states. However, it should be noted that the protocol does not generally purify the state, hence the output of the distillation  will still be a mixed state \cite{Fiurasek2010,Fiurasek2025}. Simultaneous CV entanglement distillation and purification requires more sophisticated approaches and protocols such as the one proposed in Ref.~\cite{Fiurasek2010}.

\section{Equivalent entanglement distillation scheme}

Let us first consider a single iteration of the Gaussification protocol. The original scheme that combines four single-photon  subtractions and two projections onto vacuum is depicted in Fig.~\ref{figsingle}(a). As illustrated in Fig.~\ref{figsingle}, this scheme can be converted by a sequence of transformations into the equivalent setup in Fig.~\ref{figsingle}(e). Our discussion closely follows  the simplification of scheme for generation of single-mode GKP states reported in Ref.~\cite{Rad2025}, and we apply each step of the transformation in parallel to both Alice's and Bob's side of the scheme. First we introduce additional balanced beam splitters that couple the input auxiliary vacuum modes, see Fig.~\ref{figsingle}(b). Next, we change the order of beam splitter couplings, which results in the setup depicted in Fig.~\ref{figsingle}(c). This is possible due to the following identity,
\begin{equation}
\hat{U}_{A_1A_2}\hat{V}_{A_1C_1} \hat{V}_{A_2C_2}\hat{U}_{C_1C_2}^\dagger= \hat{U}_{C_1 C_2}^\dagger \hat{V}_{A_1C_1} \hat{V}_{A_2C_2} \hat{U}_{A_1A_2}
\label{beamsplittersorder}
\end{equation}
Here $\hat{U}$ and $\hat{V}$ denote the balanced and unbalanced  beam splitter unitary operations, respectively. Unitary operation $\hat{U}$ is defined in Eq. (\ref{UBSAB}) and 
\begin{equation}
\hat{V}_{AC}=\exp[\theta(\hat{a}^\dagger \hat{c}-\hat{a} \hat{c}^\dagger)], 
\label{Vdefinition}
\end{equation}
where $\theta=\arccos \sqrt{T}$. 

In the next step we make use of the well-known fact that coupling two identical Gaussian states at beam splitters leaves them unchanged. More specifically, we have
\begin{equation}
 \hat{U}_{A_1A_2}  \hat{U}_{B_1B_2} \left(\hat{\rho}_{A_1B_1} \otimes \hat{\rho}_{A_2 B_2} \right) \hat{U}_{A_1 A_2}^\dagger  \hat{U}_{B_1B_2}^\dagger= \hat{\rho}_{A_1B_1} \otimes \hat{\rho}_{A_2B_2}
\end{equation}
where $\hat{\rho}$ is an arbitrary two-mode Gaussian state with vanishing coherent displacement. This identity implies that the two balanced beam splitters in Fig.~\ref{figsingle}(c) which couple the two copies of the input state can be removed. This results in the setup in Fig.~\ref{figsingle}(d). Finally, the partial Gaussian measurements on the second copy of the input Gaussian state converts this state into a different input two-mode state $\hat{\sigma}$ and we arrive at the final simplified setup depicted in Fig.~\ref{figsingle}(e). As can be seen in Fig.~\ref{figsingle}(d), the state $\hat{\sigma}$ is obtained from $\hat{\rho}$ by two-mode noiseless atenuation, which can be equivalently interpreted as conditional zero-photon subtraction from each mode \cite{Micuda2012,Nunn2021},
\begin{equation}
\hat{\sigma}_{AB}=\frac{1}{P}_\sigma (1-T)^{(\hat{n}_A+\hat{n}_B)/2} \hat{\rho}_{AB}  (1-T)^{(\hat{n}_A+\hat{n}_B)/2},
\end{equation}
where
\begin{equation}
P_\sigma=\mathrm{Tr}\left[ (1-T)^{\hat{n}_A+\hat{n}_B} \hat{\rho}_{AB}  \right]
\label{Psigma}
\end{equation}
is the probability of conditional preparation of state $\hat{\sigma}$  from $\hat{\rho}$ by the noiseless attenuation of each mode. The success probability of the simplified setup Fig.~\ref{figsingle}(e) is increased by factor $P_{\sigma}^{-1}$ with respect to the original scheme shown in Fig.~\ref{figsingle}(a). A simple lower bound on the probability $P_{\sigma}$ is provided by the probability of vacuum, $P_{\sigma}> \rho_{00,00}$, where $ \rho_{00,00} =\langle 00|\hat{\rho}|00\rangle$. The increase of the success probability by the elimination of the two Gaussian projections onto vacuum is typically relatively modest unless the input state is  strongly squeezed.  For pure input two-mode squeezed vacuum state (\ref{TMSV}) the state $\hat{\sigma}$ is also a pure two-mode squeezed vacuum state with reduced two-mode squeezing parameter $\nu=(1-T)\lambda$ and 
\begin{equation}
P_\sigma= \frac{1-\lambda^2}{1-(1-T)^2\lambda^2}
\end{equation}
We emphasize that the Gaussian state $\hat{\sigma}$ can be generated and distributed deterministically similarly as the original state $\hat{\rho}$, as illustrated in the final simplified setup in Fig.~\ref{figsingle}(e). The squeezing and entanglement of $\hat{\sigma}$ is weaker than that of $\hat{\rho}$. Therefore, resources that enable generation of $\hat{\rho}$ should also enable  deterministic generation and distribution of $\hat{\sigma}$, c.f. also subsequent discussion in Sec. VI.

Note that the improvement of success probability can be much higher in schemes with homodyne detections where one conditions on particular values of homodyne detector outcomes. The projection onto vacuum can be approximately implemented with  an eight-port homodyne detector that performs measurement in the overcomplete basis of coherent states, governed by POVM  $\hat{\Pi}(\alpha)=\frac{1}{\pi}|\alpha\rangle \langle \alpha|$ \cite{Eisert2007}. Projection onto vacuum can then be approximated by accepting only the outcomes $|\alpha|\leq \delta$, where $\delta$ is sufficiently small, which results in additional overhead in terms of reduced success probability \cite{Dostal2024}. Additionally, the enhancement of success probability becomes more significant for multicopy schemes, where the factor $P_{\sigma}^{-1}$ is gained for each copy of the state (except the first copy which remains in state $\hat{\rho}$). This leads to exponential improvement of success probability by factor $P_{\sigma}^{1-M}$, where $M$ is the total number of copies of the state. More details on multicopy simplified protocol can be found in Sec.~VII.

\section{Pure input states}
In this section we analyze the single iteration of entanglement distillation scheme as depicted in Fig.~\ref{figsingle} for input pure two-mode squeezed vacuum states (\ref{TMSV}).
When two copies of the input state $|\psi_{\mathrm{TMSV}}(\lambda)\rangle$ are inserted into the setup depicted in Fig.~\ref{figsingle}(a), we obtain the following conditional output state,
\begin{equation}
|\psi_{\mathrm{out}}\rangle= \sqrt{\mathcal{N}}\sum_{n=0}^\infty (n^2+3n+4) \mu^n |n,n\rangle_{AB}
\label{psiout}
\end{equation}
where $\mu=T\lambda$ is the rescaled squeezing parameter and 
\begin{equation}
\mathcal{N}=\frac{(1-\mu^2)^5}{4(4-4\mu^2+9\mu^4-4\mu^6+\mu^8)}
\label{Norm}
\end{equation}
is normalization factor. The probability of successful preparation of the state (\ref{psiout}) by the scheme in Fig.~\ref{figsingle}(a) reads
\begin{equation}
p_{S}=\frac{(1-T)^4 \lambda^4(1-\lambda^2)^2}{16\mathcal{N}}.
\label{pSoriginal}
\end{equation}

The effective transformation mapping the input Gaussian two-mode squeezed vacuum state (\ref{TMSV}) onto the output state (\ref{psiout}) can be interpreted as a combination of noiseless attenuation \cite{Micuda2012,Nunn2021} which reduces the squeezing parameter to $T\lambda$ and a transformation by quadratic operator in photon number of mode A, 
\begin{equation}
\hat{O}=\hat{n}_A^2+3\hat{n}_A+4.
\end{equation}
Note that due to the perfect photon number correlations in TMSV we could also equally consider an operator acting on mode B. The operator $\hat{O}$ approximates a noiseless quantum amplifier \cite{Ralph2009,Fiurasek2009,Zavatta2011,Neset2025} with amplitude gain  $g=2$. Indeed, as pointed out in the previouss section, for $2T\lambda<1$ the iterative Gaussification of state (\ref{TMSVsubtracted}) converges to a two-mode squeezed vacuum state with squeezing parameter $\lambda_\infty=2T\lambda$. Noiseless amplification of one part of two-mode state is one of possible mechanisms of continuous-variable entanglement concentration and distillation \cite{Ulanov2015,Seshadreesan2019,Guanzon2023}. In particular, noiseless amplification realized via single-photon catalysis was experimentally utilized in Ref. \cite{Ulanov2015} to  enhance entanglement  of two-mode squeezed state.

Let us now consider the equivalent scheme in Fig.~\ref{figsingle}(e). As already noted in Sec.~II,  the second input Gaussian  state $\hat{\sigma}$ remains pure but  its squeezing parameter is reduced to $\nu=(1-T)\lambda$. 
The sequence of a balanced beam splitter followed by  projection of each output mode onto Fock state $|1\rangle$ is equivalent to projecting the two input modes onto entangled two-photon state $\frac{1}{\sqrt{2}}(|2,0\rangle-|0,2\rangle)$. Taking into account the perfect photon number correlations exhibited by two-mode squeezed vacuum, we can conclude that the ancilla output modes C$_1$ and D$_1$   of the two unbalanced beam splitters in the scheme in Fig.~\ref{figsingle}(e) are projected onto the following non-normalized state
\begin{equation}
|\phi\rangle_{C_1D_1}=\frac{\sqrt{1-\nu^2}}{2}\left(\nu^2 |0,0\rangle+|2,2\rangle \right).
\end{equation}
It follows that the setup in Fig.~\ref{figsingle}(e) implements a coherent superposition of two-sided zero-photon subtractions and two-photon subtractions on the input state in the input modes A$_1$ and B$_1$.

Therefore, the non-normalized state generated by the simplified scheme in Fig.~\ref{figsingle}(e) can be expressed as 
\begin{equation}
\frac{\sqrt{1-\nu^2}}{2}\left[\nu^2 \hat{K}_{0,A} \otimes \hat{K}_{0,B}+\hat{K}_{2,A} \otimes \hat{K}_{2,B}\right] |\psi_{\mathrm{TMSV}}(\lambda)\rangle_{AB}.
\end{equation}
On inserting the explicit expressions (\ref{Km}) for operators $\hat{K}_m$ we get
\begin{equation}
\frac{\sqrt{1-\nu^2}}{4} (1-T)^2  T^{(\hat{n}_A+\hat{n}_B)/2}\left(2\lambda^2+\hat{a}^2\hat{b}^2\right) |\psi_{\mathrm{TMSV}}(\lambda)\rangle_{AB}.
\end{equation}
A straightforward calculation confirms that this state is exactly equal to the state (\ref{psiout}) as expected, since the two schemes in Fig.~\ref{figsingle}(a) and \ref{figsingle}(e) are fully equivalent. The only difference is that the success probability of the scheme in Fig.~\ref{figsingle}(e) is higher by the factor $1/P_\sigma$ than the success probability (\ref{pSoriginal}) of the original scheme.

The  scheme depicted in Fig.~\ref{figsingle}(e) suggests that we could treat $\nu$ as a free parameter (possibly limited by $\lambda$). We set $\nu=\kappa (1-T)\lambda$ and investigate the performance of the distillation protocol in dependence on $\kappa$. The corresponding output partially Gaussified state  can be expressed as
\begin{equation}
|\psi_{\mathrm{out}}^\prime\rangle= \sqrt{\mathcal{N}(\kappa)}\sum_{n=0}^\infty (n^2+3n+2+2\kappa^2) \mu^n |n,n\rangle_{AB},
\label{psioutprime}
\end{equation}
where
\begin{equation}
\mathcal{N}(\kappa)=\frac{1}{4}\frac{(1-\mu^2)^5}{\left[(1+\kappa^2(1-\mu^2)^2\right]^2+4\mu^2+\mu^4}.
\end{equation}
For $\kappa=1$ we recover the formulas in Eqs. (\ref{psiout}) and (\ref{Norm}).

\begin{figure}
\centerline{\includegraphics[width=0.85\linewidth]{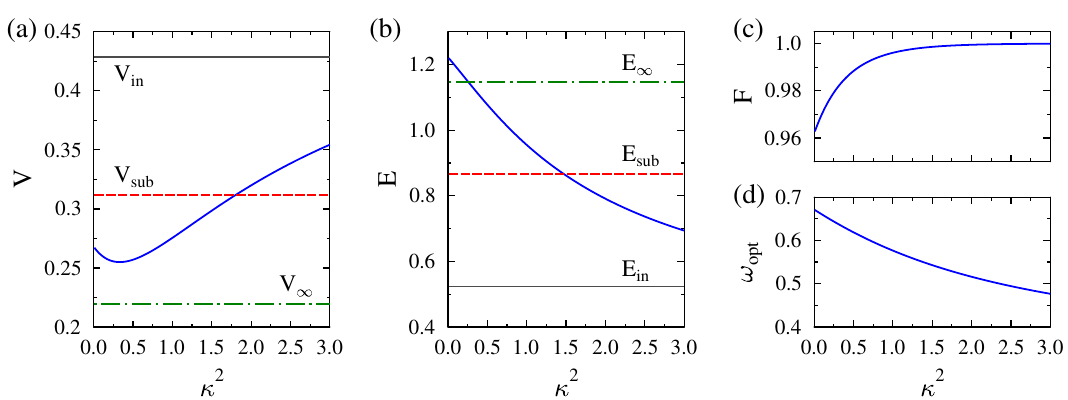}}
\caption{Performance of the simplified entanglement distillation protocol for pure input states. The squeezing variance $V$ of the distilled state $|\psi_{\mathrm{out}}^\prime\rangle$ (a), its entropy of entanglement $E$ (b), and its maximal fidelity $F$ with Gaussian two-mode squeezed vacuum state  (c)  are plotted in dependence on $\kappa^2$ (blue curves). The other parameters read $\lambda=0.4$ and $T=0.8$. The straight lines in panels (a) and (b) indicate the squeezing variance and entropy of entanglement of the  input two-mode squeezed vacuum state (in), state after local single-photon subtractions (sub), and the asymptotic TMSV state with increased squeezing parameter ($\infty$). Panel (d) shows the squeezing constant $\omega$ that maximizes the fidelity $F$ for given $\kappa^2$.}
\label{figpure}
\end{figure}

Without loss of generality, we shall assume that $\lambda$ is real and positive. The two-mode squeezed vacuum state (\ref{TMSV})  exhibits squeezing of  the difference of the amplitude  quadratures $\hat{x}_{-}=\hat{x}_A-\hat{x}_B$ and the sum of the phase quadratures $\hat{p}_{+}=\hat{p}_A+\hat{p}_B$. Note that the single-mode quadratures are defined as  $\hat{x}=\frac{1}{\sqrt{2}}(\hat{a}+\hat{a}^\dagger)$ and $\hat{p}=\frac{i}{\sqrt{2}}(\hat{a}^\dagger-\hat{a})$. For any pure state that exhibits perfect photon number correlations between modes A and B  the variances of $\hat{x}_-$ and $\hat{p}_+$ are identical and we can define a  single squeezing variance as
\begin{equation}
V= \langle(\Delta \hat{x}_A-\Delta \hat{x}_B)^2 \rangle = \langle(\Delta \hat{p}_A+\Delta \hat{p}_B)^2 \rangle .
\end{equation}
This variance is normalized such that $V=1$ for the vacuum state, and we have  $V_{\mathrm{TMSV}}=e^{-2r}=\frac{1-\lambda}{1+\lambda}$ for the Gaussian two-mode squeezed vacuum state (\ref{TMSV}). For the distilled state (\ref{psioutprime}) we obtain
\begin{equation}
V_{\mathrm{dist}}= \frac{1-\mu}{1+\mu} \frac{1-4\mu+12\mu^2-8\mu^3+5\mu^4+2\kappa^2(1-2\mu)(1-\mu^2)^2+\kappa^4(1-\mu^2)^4}{\mu^4+4\mu^2+[1+\kappa^2(1-\mu^2)^2]^2}.
\label{Vdist}
\end{equation}
The achievable squeezing depends on $\kappa$ which can be optimized to  minimize the squeezing variance $V_{\mathrm{dist}}$. The condition $\frac{\partial V_{\mathrm{dist}}}{\partial \kappa}=0$  leads to quadratic equation for $\kappa^2$, whose two roots read
\begin{equation}
\kappa^2=\frac{1}{(1-\mu^2)^4}\left(-1+2\mu-3\mu^3+3\mu^4-2\mu^6+\mu^7\pm \mu(1-\mu^2)^2\sqrt{8-8\mu+9\mu^2-4\mu^3+\mu^4} \right).
\label{kapparootV}
\end{equation}
Note that $|\kappa|<1/(1-T)$ should hold, i.e. the squeezing of the ancilla state $\hat{\sigma}$ should not exceed the squeezing of the input state $\hat{\rho}$, otherwise their roles could be interchanged. On the other hand, the parameter $\kappa^2$ can be in principle both positive and negative, because the squeezing parameter $\nu$ can be made purely imaginary by suitable phase adjustments. For comparison, we also recall the squeezing variance of the photon subtracted state (\ref{TMSVsubtracted}),
\begin{equation}
V_{\mathrm{sub}}=\frac{1-\mu}{1+\mu}\frac{1-2\mu+3\mu^2}{1+\mu^2}.
\label{Vsub}
\end{equation}
In Fig.~\ref{figpure}(a) we give an example of the dependence of the squeezing variance (\ref{Vdist}) on $\kappa^2$. For reference, the squeezing variances of the input Gaussian two-mode squeezed vacuum state (\ref{TMSV}) and the photon subtracted state (\ref{TMSVsubtracted}) are also indicated in the figure. Furthermore, as an additional benchmark, we show also the asymptotically achievable squeezing variance after full iterative Gaussification, 
\begin{equation}
V_\infty=\frac{1-2T\lambda}{1+2T\lambda}.
\end{equation}
The graph confirms that injection of auxiliary weakly squeezed two-mode squeezed vacuum state $\hat{\sigma}$ into the scheme in Fig.~\ref{figsingle}(e)  indeed helps to increase the squeezing and reduce the squeezing variance. The minimum is achieved at the positive root (\ref{kapparootV}) which is situated here at $\kappa^2 <1$. 

Quantum correlations between modes A and B can be quantified by the entropy of entanglement. For any pure bipartite state $|\psi\rangle_{AB} $ the entropy of entanglement is defined as the von Neumann entropy of reduced density matrix of mode A (or equivalently mode B), $E=-\mathrm{Tr}[\hat{\rho}_A \ln\hat{\rho_A}]$, where $\hat{\rho}_{A}=\mathrm{Tr}_B|\psi\rangle\rangle \psi|$. An example of the dependence  of entropy of entanglement of the distilled state (\ref{psioutprime}) on $\kappa^2$ is plotted in Fig.~\ref{figpure}(b). We can se that for the parameters considered  the entanglement monotonically decreases with increasing $\kappa^2$. For reference the figure indicates also the entanglement of the input two-mode squeezed vacuum state, the photon subtracted state, and the asymptotically distilled two-mode squeezed vacuum  state with increased squeezing parameter $2T\lambda$.

To obtain further insight into the role of $\kappa$, we consider fidelity of the distilled state (\ref{psioutprime}) with two-mode squeezed vacuum state with squeezing parameter $\omega$,
\begin{equation}
F=\frac{(1-\mu^2)^5(1-\omega^2)\left[1+\kappa^2(1-\mu\omega)^2\right]^2}{(1-\mu\omega)^6\left(\mu^4+4\mu^2+[1+\kappa^2(1-\mu^2)^2]^2\right)}.
\label{Fomega}
\end{equation}
The squeezing parameter $\omega$ that maximizes the fidelity (\ref{Fomega}) can be found by solving the equation $\frac{\partial F}{\partial \omega}=0$ which yields a cubic equation for $\omega$,
\begin{equation}
3\mu+\kappa^2\mu-(1+\kappa^2+2\kappa^2\mu^2)\omega+[\kappa^2\mu(\mu^2+2)-2\mu]\omega^2-\kappa^2\mu^2\omega^3=0.
\end{equation}
In  Fig.~\ref{figpure}(c) we plot the maximally achievable fidelity (\ref{Fomega}) in dependence on $\kappa$ and Fig.~\ref{figpure}(d) shows the corresponding squeezing parameter $\omega$  that maximizes $F$ for given $\mu$ and $\kappa$.
We can see that with increasing $\kappa^2$ the fidelity increases and approaches $1$ while $\omega$ decreases. This can be understood by observing that for very  large $\kappa$ the distilled state (\ref{psioutprime}) approaches the two-mode squeezed vacuum state with squeezing parameter $\mu$. We can also observe that for $\kappa^2 \lesssim 1$  the fidelity rapidly increases with $\kappa$. By tuning $\kappa$, one may therefore sacrify a bit of the two-mode squeezing to get a state which is closer to a Gaussian state.

\begin{figure}
\centerline{\includegraphics[width=0.4\linewidth]{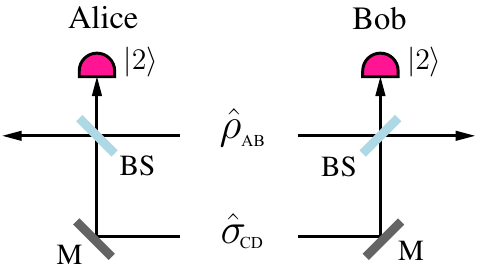}}
\caption{Entanglement distillation by two-mode generalized photon subtraction \cite{DellAnno2013}. Two photons are conditionally subtracted from each mode of the input state $\hat{\rho}_{AB}$.  Auxiliary Gaussian two-mode squeezed state $\hat{\sigma}$ is injected into the auxiliary input ports of the unbalanced beam splitters BS that serve for photon subtraction. Mirrors M direct the auxiliary optical beams into the input ports of BS.}
\label{figgeneralized}
\end{figure}

\section{Comparison with bipartite generalized photon subtraction}

In this section we compare the simplified entanglement distillation scheme depicted in Fig.~\ref{figsingle}(e) with a conceptually similar scheme that was proposed  in Ref. \cite{DellAnno2013}. This scheme is depicted in Fig.~\ref{figgeneralized} and it  can be interpreted as a nonlocal two-mode version of the generalized photon subtraction protocol \cite{Takase2021,Tomoda2024,Takase2024} that has been recently suggested for generation of approximate GKP states. In the scheme shown in Fig.~\ref{figgeneralized}, two photons are subtracted from each mode A and B of a two-mode quantum state $\hat{\rho}_{\mathrm{AB}}$, but an ancilla Gaussian two-mode state $\hat{\sigma}_{\mathrm{CD}}$ is injected into the auxiliary input ports of the beam splitters. Therefore, the detected heralding photons can come either from state $\hat{\rho}$ or $\hat{\sigma}$. 

To obtain insight into the working of the scheme depicted in Fig.~\ref{figgeneralized}, we again consider pure input two-mode squeezed vacuum states.  It is convenient to  represent these  input  states as functions of creation operators acting on vacuum states, which is closely related to the Bargmann representation of quantum states of bosonic systems \cite{Vourdas2006,Motamedi2025},
\begin{equation}
|\Psi_{\mathrm{in}}\rangle=|\psi_{\mathrm{TMSV}}(\lambda)\rangle_{AB}|\psi_{\mathrm{TMSV}}(\nu)\rangle_{CD}=\sqrt{(1-\lambda^2)(1-\nu^2)} e^{\lambda \hat{a}^\dagger \hat{b}^\dagger+\nu \hat{c}^\dagger \hat{d}^\dagger} |\mathrm{vac}\rangle.
\label{TMSVBargmann}
\end{equation}
Here $|\mathrm{vac}\rangle$ denotes the multimode vacuum state. The conditionally generated state of modes A and B prepared  by the scheme in Fig.~\ref{figgeneralized} can be expressed as 
\begin{equation}
|\psi_{\mathrm{cond}}\rangle_{AB}=_{CD}\langle 2,2| \hat{V}_{AC}\hat{V}_{BD} |\Psi_{\mathrm{in}}\rangle,
\label{psicond}
\end{equation}
where $\hat{V}$ is a unitary operation describing interference at an unbalanced beam splitter with intensity transmittance $T$, c.f. Eq. (\ref{Vdefinition}).
Making use of the representation (\ref{TMSVBargmann}) and taking into account the vacuum stability condition $\hat{V}_{AC}\hat{V}_{BD}|\mathrm{vac}\rangle=|\mathrm{vac}\rangle$ we get
\begin{equation}
|\psi_{\mathrm{cond}}\rangle_{AB}=_{CD}\langle 2,2|  \exp\left[\tilde{\lambda} \hat{a}^\dagger \hat{b}^\dagger+\tilde{\nu} \hat{c}^\dagger \hat{d}^\dagger +\sqrt{(1-T)T}(\lambda-\nu)(\hat{a}^\dagger \hat{d}^\dagger+\hat{b}^\dagger \hat{c}^\dagger)\right] |\mathrm{vac}\rangle,
\end{equation}
where $\tilde{\lambda}=T\lambda+(1-T)\nu$ and $\tilde{\nu}=T\nu+(1-T)\lambda$. Next we expand the terms that contain the creation operators $\hat{c}^\dagger$ or $\hat{d}^\dagger$ in Taylor series. After a straightforward algebra we obtain
\begin{equation}
|\psi_{\mathrm{cond}}\rangle_{AB}=\sqrt{(1-\lambda^2)(1-\nu^2)}\sum_{n=0}^\infty \left[ \tilde{\nu}^2\tilde{\lambda}^2+2\tilde{\nu}\tilde{\lambda}T(1-T)(\lambda-\nu)^2 n+\frac{1}{2}T^2(1-T)^2(\lambda-\nu)^4 n(n-1)\right] \tilde{\lambda}^{n-2}|n,n\rangle.
\end{equation}
The structure of this state resembles that of the state (\ref{psiout}). The squeezing parameter is modified to $\tilde{\lambda}$ and the two-mode Fock-state amplitudes are modulated by a quadratic polynomial in $n$, which we formally denote as $n^2+d_1n+d_0.$ The coefficients of this polynomial can be partly tuned and adjusted by changing $T$ and $\nu$. However, it should be noted that the schemes in Fig.~\ref{figsingle} and Fig.~\ref{figgeneralized} are not fully equivalent. In particular, with the setup depicted in Fig.~\ref{figgeneralized} it is not possible to implement the quadratic modulation described by polynomial  $n^2+3n+4$, because when we set $d_1=3$ we find that this implies $\nu=0$ and $d_0=2$.

\section{Mixed input states}

\begin{figure}
\centerline{\includegraphics[width=\linewidth]{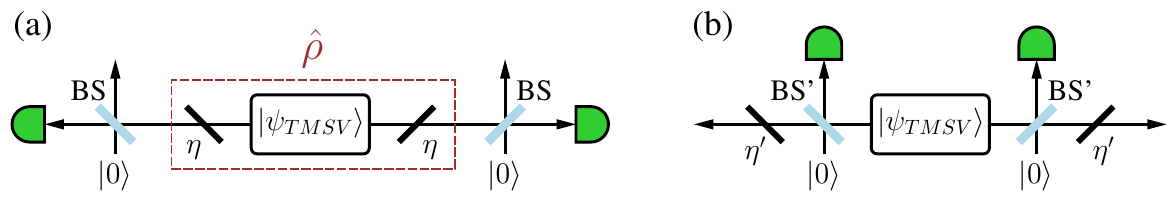}}
\caption{Preparation of Gaussian state $\hat{\sigma}$ from input mixed Gaussian state $\hat{\rho}$. (a) Scheme extracted from Fig.~\ref{figsingle}(d). (b) An equivalent scheme, where the order of zero-photon subtractions and ordinary attenuations is reversed. Beam splitters labeled by $\eta$ denote ordinary loss channels, and the green detectors perform projection onto vacuum. See main text for explicit relations between the parameters of the two schemes. }
\label{figsigmaequivalence}
\end{figure}

Let us now consider mixed input states $\hat{\rho}$. It is convenient to calculate the output partially distilled state directly for the simplified setup in Fig.~\ref{figsingle}(e). We obtain
\begin{equation}
\hat{\rho}_{\mathrm{dist}}=\frac{1}{4}\sum_{j=0}^1 \sum_{k=0}^1 \sum_{l=0}^1 \sum_{m=0}^1(-1)^{j+k+l+m} \hat{K}_{2j,A} \hat{K}_{2k,B} \hat{\rho}_{AB}  \hat{K}_{2l,A}^\dagger \hat{K}_{2m,B}^\dagger  \sigma_{\bar{j}\bar{k},\bar{l}\bar{m}},
\end{equation}
where $\sigma_{ab,cd}=\langle ab|\hat{\sigma}|cd\rangle$ denotes the density matrix elements of $\hat{\sigma}$ in Fock basis and $\bar{x}=2-2x$. Let us assume that the input mixed state $\hat{\rho}$ is obtained from  pure-mode squeezed vacuum state (\ref{TMSV}) with squeezing parameter  $\lambda$ whose both modes are transmitted through a lossy quantum channel with transmittance $\eta$. The Gaussian state $\hat{\rho}$ is then fully characterized by  covariance matrix $\gamma_{\rho}=\eta\gamma_{\mathrm{TMSV}}(\lambda)+(1-\eta)I$, where $\gamma_{\mathrm{TMSV}}$ denotes the covariance matrix of pure two-mode squeezed vacuum state (\ref{TMSV}) and $I$ is the covariance matrix of vacuum state, which is equal to the identity matrix. The state $\hat{\sigma}$ which is obtained from the state $\hat{\rho}$ by local noiseless attenuations can be represented in a similar manner, as a two-mode squeezed vacuum state with squeezing parameter $\nu$ whose both modes are transmitted over a lossy channel with transmittance $\eta^\prime$. The parameters $\eta^\prime$ and $\nu$ can be expressed as follows, 
\begin{equation}
\nu=\lambda(1-\eta T),\qquad \eta^\prime=\eta \frac{1-T}{1-\eta T}.
\label{etanuprime}
\end{equation}
If Fig.~\ref{figsigmaequivalence} we illustrate two equivalent schemes for preparation of state $\hat{\sigma}$ which are instrumental for derivation of the formulas (\ref{etanuprime}) \cite{Fiurasek2025}. 
Beam splitters BS have intensity transmittance $T$, while the beam splitters BS$^\prime$ in Fig.~\ref{figsigmaequivalence}(b) have transmittance $T^{\prime}=1-\eta T$. This ensures that the same amount of signal impinges on the detectors in both schemes. Furthermore, the relation between $\eta$ and $\eta^\prime$ can be derived from the requirement that the same portion of the input signal reaches the signal output modes, $\eta(1-T)=\eta^\prime T^\prime$.

Observe that $\eta^{\prime} \leq \eta$. This is important, because it shows that the lossy channels that are used to distribute state $\hat{\rho}$ can be also used to distribute state $\hat{\sigma}$. There is even some space for improvement, because we do not have to impose losses higher than $\eta$. The state $\hat{\sigma}$ can be equivalently expressed as two-mode squeezed thermal state,
\begin{equation}
\hat{\sigma}=\hat{S}_{CD}(s)  (\hat{\tau}_{C}\otimes\hat{\tau}_{D})\hat{S}^\dagger_{CD}(s),
\label{sigmathermal}
\end{equation}
where $\hat{S}_{CD}(s)=\exp(s\hat{c}^\dagger \hat{d}^\dagger-s^\ast \hat{c}\hat{d})$ is the two-mode squeezing operator and 
\begin{equation}
\hat{\tau}=\sum_{n=0}^\infty \frac{1}{\bar{n}+1}\left( \frac{\bar{n}}{\bar{n}+1}\right)^n |n\rangle \langle n|
\end{equation}
is a single-mode thermal state with mean number of thermal photons equal to $\bar{n}$. Specifically for the state $\hat{\sigma}$ with parameters (\ref{etanuprime}) we obtain
\begin{equation}
\tanh (2s)=\frac{2\eta\lambda(1-T)}{1-\lambda^2[1-2\eta+\eta^2(2-T)T]}, \qquad \bar{n}=\frac{1}{2} \left[\sqrt{\frac{1-\lambda^2[1-\eta(2-T)]^2}{1-\lambda^2(1-\eta T)^2}}-1\right].
\label{nbarsigma}
\end{equation}
It follows from Eq. (\ref{sigmathermal}) that $\hat{\sigma}$ is a mixture of two-mode squeezed Fock states. The dominant term is the two-mode squeezed vacuum term, which occurs with probability $1/(\bar{n}+1)^2$. We are mostly interested in the parameter range where the auxiliary stat eis close to a pure state. Assuming $\bar{n}\ll 1$ we can approximate Eq. (\ref{nbarsigma}) as
\begin{equation}
\bar{n}\approx \frac{\eta(1-\eta)\lambda^2 (1-T)}{1-\lambda^2(1-\eta T)^2},
\end{equation}
which more clearly illustrates the dependence of $\bar{n}$ on the various parameters of the protocol. 

\begin{figure}
\includegraphics[width=\linewidth]{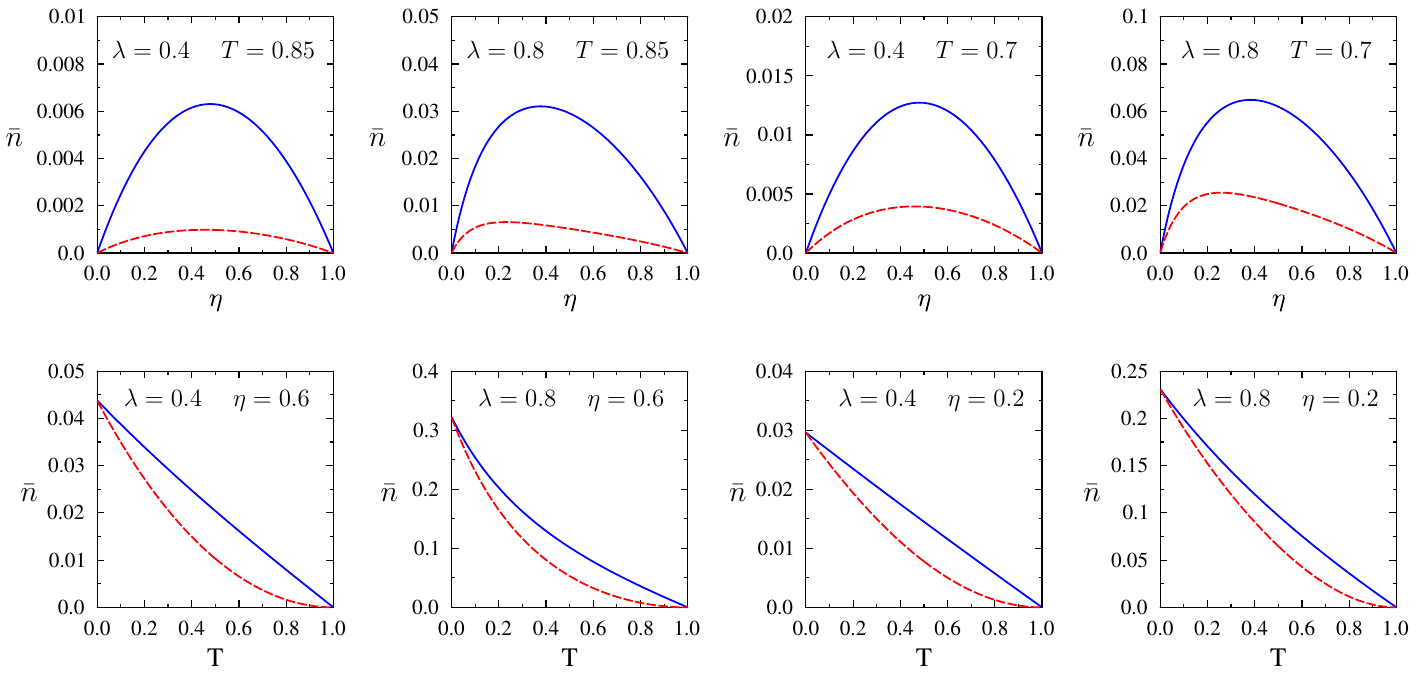}
\caption{Dependence of the mean numbers of thermal photons $\bar{n}$ (solid blue lines) and $\bar{n}^\prime$ (red dashed lines) on $\eta$ (or $T$) is plotted for two different values of $\lambda$ and selected values of $T$ (or $\eta$).}
\label{fignbar}
\end{figure}

By preparing two-mode squeezed vacuum state with suitable squeezing and distributing it over lossy channels with transmittance $\eta$ we can generate a state $\hat{\sigma}^\prime$ with the same squeezing as in Eq. (\ref{sigmathermal}) but with reduced thermal noise,
\begin{equation}
\bar{n}^\prime=\frac{1}{2}\left[\frac{2\eta\nu^\prime}{(1-\nu^{\prime 2})\sinh(2s)}-1\right], \qquad \nu^\prime=\frac{\sqrt{\eta^2+(1-2\eta)\tanh^2(2s)}-\eta}{\tanh(2s)(1-2\eta)}.
\end{equation}
The dependence of the mean numbers of thermal photons $\bar{n}$ and $\bar{n}^\prime$ on $T$ and $\eta$ is illustrated in Fig.~\ref{fignbar} for two different values of $\lambda$. 

\begin{figure}
\includegraphics[width=\linewidth]{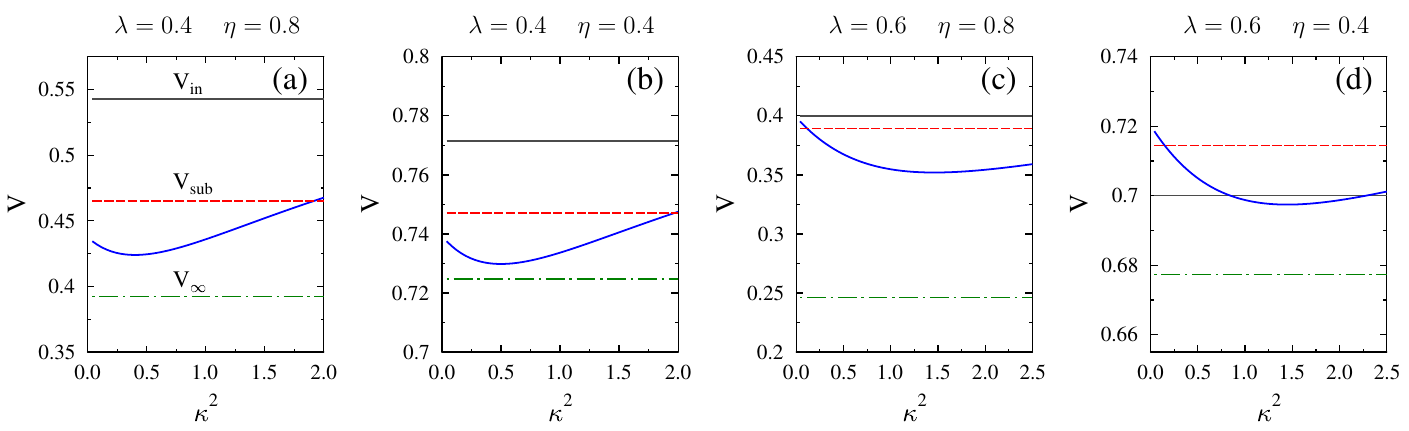}
\caption{Distillation of mixed states by the scheme in Fig.~\ref{figsingle}(e). Dependence of the squeezing variance $V$ of the distilled state on $\kappa^2$ is plotted for several different combinations of $\lambda$ and $\eta$, and $T=0.8$ (blue solid curves). Note that $\kappa$ characterizes the squeezing of the auxiliary mixed state $\hat{\sigma}$, as specified in the main text. For reference and comparison, each panel also displays the  corresponding squeezing variance $V_{\mathrm{in}}$ of the input Gaussian state $\hat{\rho}$ (gray solid line) , squeezing variance $V_{\mathrm{sub}}$ of the mixed photon subtracted state (red dashed line), and variance $V_{\infty}$  of the asymptotic Gaussian state that would be distilled from the photon subtracted state by iterative heralded Gaussification (green dot-dashed line).}
\label{figmixed}
\end{figure}

Similarly as for the pure states, we can vary the squeezing parameter of the state $\hat{\sigma}$ in some range. To be specific, we consider that $\hat{\sigma}$ is obtained by distributing  a pure two-mode squeezed vacuum state with squeezing parameter $\kappa \nu^\prime$ over two symmetric lossy channels with transmittance $\eta$, c.f. Fig.~\ref{figsigmaequivalence}(b). In Fig.~\ref{figmixed} we plot the squeezing variance of the distilled state as function of $\kappa^2$ for four different combinations of parameters $\eta$ and $\lambda$. For reference,  each graph also indicates the squeezing variance of the input mixed Gaussian state 
\begin{equation}
V_{\mathrm{in}}=\frac{1-\lambda}{1+\lambda} \eta+1-\eta,
\end{equation}
and the variance of the mixed photon subtracted state
\begin{equation}
V_{\mathrm{sub}}=\frac{1-\tilde{\mu}}{1+\tilde{\mu}}\frac{1-2\tilde{\mu}+3\tilde{\mu}^2}{1+\tilde{\mu}^2}\tilde{\eta}+1-\tilde{\eta}, \qquad \tilde{\mu}=(1-\eta+\eta T)\lambda, \qquad \tilde{\eta}= \frac{\eta T}{1-\eta(1-T)}.
\label{Vsubmixed}
\end{equation}
This formula for $V_{\mathrm{sub}}$ can be derived by utilizing the equivalence of the two setups depicted in Fig.~\ref{figsigmaequivalence}. Namely, the photon subtraction can be effectively performed on a pure two-mode squeezed vacuum state and this state is then propagated through lossy channels. Using the analytical formula for squeezing variance of pure photon-subtracted two-mode squeezed vacuum  state (\ref{Vsub}) we can straightforwardly derive Eq. (\ref{Vsubmixed}).

Moreover, we also plot in Fig.~\ref{figmixed} the squeezing variance $V_{\infty}$ of a mixed Gaussian state that is asymptotically distilled from the photon subtracted state by the iterative Gaussification protocol depicted in in Fig.~\ref{figsingle}(f). Analytical formula for the  covariance matrix of the asymptotically distilled Gaussian state was derived by Eisert \emph{et al.} in Ref. \cite{Eisert2004}.
With the help of these results we obtain
\begin{equation}
V_{\infty}=
\frac{1+(1-\eta)\lambda[1+(1-\eta)\lambda(1+(1-\eta)\lambda)]-2\eta T\lambda[1+\lambda(-1+\eta+(1-\eta)^2 \lambda)]}{1+(1-\eta)\lambda[1+(1-\eta)\lambda(1+(1-\eta)\lambda)]+2\eta T\lambda[1+\lambda(-1+\eta+(1-\eta)^2 \lambda)]}.
\end{equation}
This formula is valid provided that the Gaussification protocol converges. 

Several conclusions can be drawn from Fig.~\ref{figmixed}. The squeezing of $\hat{\sigma}$ can be optimized to minimize the squeezing variance of the distilled state, similarly as for pure states. The asymptotically distillable squeezing is limited by losses. A simple bound is provided by $V \geq 1-\eta$. This bound can be further refined by considering the two equivalent schemes in Fig.~\ref{figsigmaequivalence} and applying this to state $\hat{\rho}$ in Fig.~\ref{figsingle}(e). This yields a stricter bound 
$V \geq 1-\tilde{\eta}$, where $\tilde{\eta}$ is specified in Eq. (\ref{Vsubmixed}). It can be seen in Fig.~\ref{figmixed} that for higher $\lambda$  the photon subtraction by itself does not necessarily  increase the squeezing, and $V_{\mathrm{sub}}>V_{\mathrm{in}}$ can hold. Nevertheless even in such cases the multicoppy distillation can be beneficial with $V_{\infty}< V_{\mathrm{in}}$. By suitably tuning $\kappa$, we can get $V<V_{\mathrm{in}}$ already with the simplified two-copy protocol.

\section{Simplified multicopy scheme}

\begin{figure*}
\includegraphics[width=0.85\linewidth]{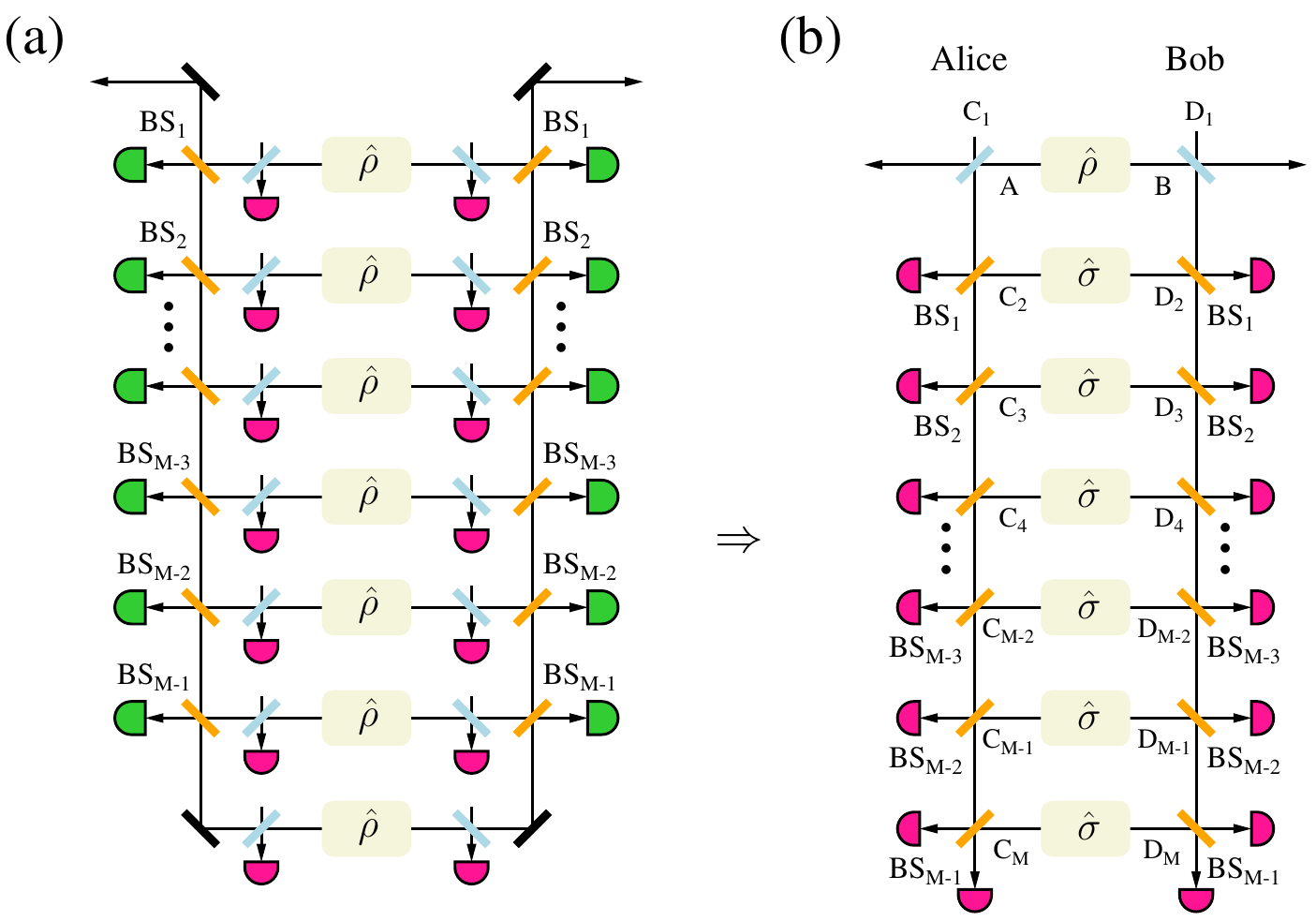}
\caption{Simplification of multicopy continuous-variable entanglement distillation scheme. (a) Original protocol. Each copy of the state $\hat{\rho}$ is first de-Gaussified by local single-photon subtractions and $M$ copies of the de-Gaussified state are Gaussified by interference in two arrays of beam splitters followed by projection af all auxiliary output modes onto vacuum. (b) Equivalent simplified scheme that involves only non-Gaussian measurements - projections onto single-photon states. The Gaussian states $\hat{\sigma}$ are obtained from the input states $\hat{\rho}$ by two-mode noiseless attenuation, as illustrated in Fig.~5 and discussed in the main text.  }
\label{figmultiple}
\end{figure*}

In this section we present simplification of a general multicopy entanglement distillation scheme. The original $M$-copy  distillation scheme is illustrated in Fig.~\ref{figmultiple}(a). $M$ copies of Gaussian state $\hat{\rho}$ are distributed to Alice and Bob. Both Alice and Bob locally apply single-photon subtractions to their modes, which de-Gaussifies the input states. Subsequently, Alice and Bob interfere their $M$ modes in $M$-mode interferometers each formed by a sequence of $M-1$ beam splitters BS$_j$ with transmittances $T_j=(M-j)/(M-j+1)$.  An important property of the beam splitter network is that the signal output mode is a balanced combination of the $M$ input modes, hence all input modes contribute equally to the output. All the remaining $M-1$ modes are projected onto vacuum state. Since the vacuum state is invariant with respect to passive Gausisan unitary transformations, any interferometric coupling of the measured modes on Alice's (or Bob's) side does not affect the scheme. This ensures, in particular, that for $M=2^N$ the considered scheme is fully equivalent to an iterative scheme where pairs of copies of the state are repeatedly Gaussified according to Eq. (\ref{Gaussificationmap}).

The $M$-copy protocol can be converted into the equivalent scheme depicted in Fig.~\ref{figmultiple}(b). The simplification proceeds in the same way as for the two-copy scheme in Fig.~\ref{figsingle}  \cite{Rad2025}. First, two arrays of auxiliary beam splitters are introduced to couple the auxilairy vacuum modes utilized for the photon subtractions.  Second, the order of the interferometric couplings is exchanged in a similar manner as in Fig.~\ref{figsingle}(c), just now it is applied to the whole beam-splitter arrays. Subsequently, the invariance of the $M$-copy Gaussian state $\hat{\rho}^{\otimes M}$ under symmetric interferometric couplings on Alice's and Bob's side is exploited to remove the innermost layers of interferometers. Finally, the action of Gaussian measurements is incorporated into the preparation of Gaussian states $\hat{\sigma}$.

Let us now analyze the resulting simplified scheme depicted in Fig.~\ref{figmultiple}(b). We shall consider pure input states for simplicity, and the goal is to understand how  Gaussification emerges in this scheme without any Gaussian measurements. Modes A and B are initially prepared in the two-mode squeezed vacuum state with squeezing parameter $\lambda$. Modes C$_1$ and D$_1$ are initially in vacuum state. Each pair of modes C$_j$D$_j$ for $j \geq 2$ is prepared in pure two-mode squeezed vacuum state with squeezing parameter $\nu=(1-T)\lambda$. 
This input multimode pure Gaussian state can be conveniently expressed as 
\begin{equation}
|\Psi_{\mathrm{in}}\rangle=\sqrt{\mathcal{M}} \exp\left( \lambda \hat{a}^\dagger\hat{b}^\dagger +\nu\sum_{j=2}^M \hat{c}_j^\dagger \hat{d}_j^\dagger\right)|\mathrm{vac}\rangle,
\end{equation}
where $|\mathrm{vac}\rangle$ denotes the multimode vacuum state and
\begin{equation}
\mathcal{M}=(1-\lambda^2)(1-\nu^2)^{M-1}.
\end{equation}
Modes A and C$_1$ (B and D$_1$) are coupled by beam splitters with transmittance $T$ which results in a state
\begin{widetext}
\begin{equation}
|\Psi\rangle=\sqrt{\mathcal{M}} \exp\left( \mu\hat{a}^\dagger\hat{b}^\dagger+\sqrt{\mu\nu}(\hat{a}^\dagger \hat{d}_1^\dagger+\hat{b}^\dagger\hat{c}_1^\dagger) +\nu\sum_{j=1}^M \hat{c}_j^\dagger \hat{d}_j^\dagger\right)|\mathrm{vac}\rangle.
\label{PsiMmode}
\end{equation}
Recall that $\mu=T\lambda$.
Subsequently, modes C$_j$ (D$_j$) are mutually coupled at two identical $M$-port interferometers described by a real coupling matrix $V$,
\begin{equation}
\hat{c}_j^\dagger \rightarrow \sum_{k}V_{jk} \hat{c}_k^\dagger, \qquad \hat{d}_j^\dagger \rightarrow \sum_{l}V_{jl} \hat{d}_l^\dagger.
\end{equation}
It holds that $V^TV=I$ where $I$ stands for the identity matrix. Moreover, $V_{1k}=\frac{1}{\sqrt{M}}$ for all $j$. This latter condition is equivalent to the condition that in the original Gaussification scheme all input modes contribute equally to the output mode.  The interferometric coupling $V$ transforms the state (\ref{PsiMmode}) into
\begin{equation}
|\Psi_2\rangle=\sqrt{\mathcal{M}} \exp\left( \mu\hat{a}^\dagger\hat{b}^\dagger+\sqrt{\frac{\mu\nu}{M}}\sum_{j=1}^M(\hat{a}^\dagger \hat{d}_j^\dagger+\hat{b}^\dagger\hat{c}_j^\dagger) +\nu\sum_{j=1}^M \hat{c}_j^\dagger \hat{d}_j^\dagger\right)|\mathrm{vac}\rangle.
\end{equation}
It is instructive to rewrite this state as
\begin{equation}
|\Psi_2\rangle=\sqrt{\mathcal{M}} \exp\left( \mu\hat{a}^\dagger\hat{b}^\dagger\right) \prod_{j=1}^M \exp\left[\sqrt{\frac{\mu\nu}{M}}(\hat{a}^\dagger \hat{d}_j^\dagger+\hat{b}^\dagger\hat{c}_j^\dagger) +\nu \hat{c}_j^\dagger \hat{d}_j^\dagger\right]|\mathrm{vac}\rangle.
\label{Psi2}
\end{equation}
Conditioning on detection of a single photon in each of the modes C$_j$ and D$_j$ finally converts the state (\ref{Psi2}) to output single-mode state
\begin{equation}
|\Psi_{\mathrm{out}}\rangle_{AB}=\sqrt{\mathcal{M}} \nu^M \left(1+\frac{\mu \hat{a}^\dagger \hat{b}^\dagger}{M}\right)^M \exp\left( \mu\hat{a}^\dagger\hat{b}^\dagger\right) |\mathrm{vac}\rangle.
\end{equation}
Each pair of modes C$_j$D$_j$ gives rise to one factor $1+\frac{\mu}{M} \hat{a}^\dagger \hat{b}^\dagger$. This can be  shown by expanding the exponentials in Eq. (\ref{Psi2}) into Taylor series and keeping only terms proportional to $\hat{c}_j^\dagger d_j^\dagger$ which generate single-photon states in the measured modes. 
In the limit of large $M$ we have $(1+\mu\hat{a}^\dagger \hat{b}^\dagger/M)^M \rightarrow \exp(\mu \hat{a}^\dagger \hat{b}^\dagger)$, hence we get $|\Psi_{\mathrm{out}}\rangle\propto \exp(2\mu \hat{a}^\dagger \hat{b}^\dagger)|\mathrm{vac}\rangle$ which is a Gaussian two-mode squeezed vacuum state with squeezing parameter $2\mu=2T\lambda$.

\section{Discussion and Conclusions}

In summary, we have derived a simplified scheme for multicopy continuous variable entanglement distillation, where all Gaussian measurements are eliminated. The resulting scheme thus contains only projections on single-photon states which originate from the de-Gaussification of the input Gaussian states by single-photon subtractions. The resulting entanglement distillation scheme can be interpreted as an instance of bipartite Gaussian boson sampling, where $M$  two-mode Gaussian states are injected into suitably designed multimode interferometers on Alice's and Bob's side and all modes except the two output modes are projected onto the single-photon state. 

The simplified entanglement distillation scheme closely resembles the simplified scheme for generation of single mode GKP state discussed in Ref. \cite{Rad2025} and it essentially represents a bipartite version of that single-mode protocol. Similarly as in Ref. \cite{Rad2025}, the simplified setup offers several degrees of freedom that can be optimized to maximize the extracted two-mode squeezing or other relevant property of the distilled state. In particular, this includes the choice of the auxiliary Gaussian state $\hat{\sigma}$ and the photon-subtraction transmittance $T$.  

The performed simplification  reduces the complexity of the scheme  and it also increases the success probability of the protocol. However, the protocol  requires that 
all single-photon detections succeed simultaneously, which may limit the practical scaling of the scheme to larger $M$.  To further improve the success probability, schemes involving quantum memories can be utilized \cite{Cotte2022,Simon2024}. However, such schemes do not fall within the class of setups considered in the present work, because the storage and release from the memory is controlled by outcomes of measurements in other parts of the setup, hence the memory-based approaches involve feedforward. Note that several adaptive schemes with feedforward/feedback have been recently proposed to increase the success probability of preparation of the single-mode GKP states or other interesting non-Gaussian states \cite{Weigand2024,Pizziment2024,Crescimanna2025}.

Our work provides new interesting  insights into the heralded Gaussification protocol. Specifically,  muticopy CV entanglement distillation can converge to a Gaussian state  even without heralding 
Gaussian measurements. The mechanism bears  similarities to approximate physical implementations of heralded noiseless quantum amplifier by conditional photon additions and subtractions \cite{Fiurasek2009,Zavatta2011,Neset2025}, where the unbounded operator $g^{\hat{n}}$ with $g>1$ is approximated by a polynomial in $\hat{n}$ multiplied by exponentially damping factor $T^{\hat{n}/2}$ to preserve physicality. Here, similarly,  the Gaussian operator $\exp(\mu \hat{a}^\dagger \hat{b}^\dagger)$ is approximated by a polynomial in $\hat{a}^\dagger \hat{b}^\dagger$ while the input two-mode squeezed vacuum state is also effectively multiplied by an exponentially damping term  which attenuates the initial squeezing from $\lambda$ to $\mu=T\lambda$.

In our analysis we have for simplicity assumed perfect detectors that can count the photons and distinguish the single-photon state $|1\rangle$ from other Fock states. Recent tremendous technological advances have brought superconducting single-photon detectors with photon number resolving capacity and detector efficiency exceeding $99\%$ \cite{Larsen2025}.  Effective losses in the experiment can also arise directly  in the source of the squeezed states or during the state transmission from the source to the detector, and the latter are particularly relevant in quantum communication and entanglement distribution. Overall, the recent experimental breakthrough in generation of optical GKP states \cite{Larsen2025} shows that multimode multicopy experiments with optical squeezed states and heralding on counts of photons are experimentally feasible. 

 Going beyond entanglement distillation, the simplified setups depicted in Figs.~\ref{figsingle} and ~\ref{figmultiple} can be also operated in a regime where the Gaussification does not converge, which could be utilized for engineering of highly non-classical and quantum non-Gaussian entangled tow-mode states, similarly as in breeding of single-mode cat states \cite{Sychev2017} or GKP states \cite{Konno2024}. However, investigation of such possibilities is beyond the scope of the present study and we leave it to future work.

\end{widetext}

\begin{acknowledgments}
This work was supported by Palacký University under Project No. IGA-PrF-2025-010.
\end{acknowledgments}

\end{document}